\documentclass[letterpaper,twocolumn,10pt, hyphens]{article}
\usepackage{usenix2024_SOUPS}
\usepackage[square,numbers]{natbib}
\usepackage{url}

\usepackage{tikz}
\usepackage{amsmath}
\usepackage{booktabs}
\usepackage[table,xcdraw]{xcolor}
\usepackage{xspace}
\usepackage{xcolor}
\usepackage{hyperref}
\hypersetup{hidelinks}
\usepackage{tcolorbox}
\usepackage{titlesec}
\titlespacing*{\paragraph}{0pt}{1ex plus 0.2ex minus 0.1ex}{1em}
\usepackage{balance} 
 
\newcommand{\techemph}[1]{\textit{\textcolor{teal}{#1}}}

\newcommand{\numpapers}{29\xspace}

\newcommand{\cellAligns}[1]{\cellcolor[HTML]{9AFF99}{#1}}
\newcommand{\cellConflicts}[1]{\cellcolor[HTML]{FFCCC9}{#1}}

\newcommand{\simplicityandnecessity}{P1: SIMPLICITY-AND-NECESSITY\xspace}
\newcommand{\safedefaults}{P2: SAFE-DEFAULTS\xspace}
\newcommand{\opendesign}{P3: OPEN-DESIGN\xspace}
\newcommand{\completemediation}{P4: COMPLETE-MEDIATION\xspace}
\newcommand{\isolatedcompartments}{P5: ISOLATED-COMPARTMENTS\xspace}
\newcommand{\leastprivilege}{P6: LEAST-PRIVILEGE\xspace}
\newcommand{\modulardesign}{P7: MODULAR-DESIGN\xspace}
\newcommand{\smalltrustedbases}{P8: SMALL-TRUSTED-BASES\xspace}
\newcommand{\timetestedtools}{P9: TIME-TESTED-TOOLS\xspace}
\newcommand{\leastsurprise}{P10: LEAST-SURPRISE\xspace}
\newcommand{\userbuyin}{P11: USER-BUY-IN\xspace}
\newcommand{\sufficientworkfactor}{P12: SUFFICIENT-WORK-FACTOR\xspace}
\newcommand{\defenseindepth}{P13: DEFENSE-IN-DEPTH\xspace}

\newcommand{\evidenceproduction}{P14: EVIDENCE-PRODUCTION\xspace}
\newcommand{\datatypevalidation}{P15: DATATYPE-VALIDATION\xspace}
\newcommand{\remnantremoval}{P16: REMNANT-REMOVAL\xspace}
\newcommand{\trustanchorjustification}{P17: TRUST-ANCHOR-JUSTIFICATION\xspace}
\newcommand{\idependentconfirmation}{P18: INDEPENDENT-CONFIRMATION\xspace}
\newcommand{\requestresponseintegrity}{P19: REQUEST-RESPONSE-INTEGRITY\xspace}
\newcommand{\reluctantallocation}{P20: RELUCTANT-ALLOCATION\xspace}
\newcommand{\securitybydesign}{HP1: SECURITY-BY-DESIGN\xspace}
\newcommand{\designforevolution}{HP2: DESIGN-FOR-EVOLUTION\xspace}

\newcommand{\simplifyandreduce}{\textit{Simplify and Reduce}\xspace}
\newcommand{\durabilityandlongevity}{\textit{Durability and Longevity}\xspace}
\newcommand{\collaboration}{\textit{Compatibility and Openness}\xspace}
\newcommand{\adaptandaugment}{\textit{Adapt and Augment}\xspace}
\newcommand{\efficiency}{\textit{Efficiency}\xspace}
\newcommand{\context}{\textit{Context and Stakeholders}\xspace}

\newcommand{\sususerbuyin}{\textit{User Buy-In\xspace}}
\newcommand{\usability}{\textit{Usability}\xspace}
\newcommand{\repairandmaintain}{\textit{Repair and Maintain}\xspace}

\newcommand{\reuse}{\textit{Reuse}\xspace}
\newcommand{\recycle}{\textit{Recycle}\xspace}
\newcommand{\degrade}{\textit{Degrade Gracefully}\xspace}

\begin{document}

\date{}

\title{SoK: Is Sustainable the New Usable? \\ \textit{Debunking The Myth of Fundamental Incompatibility Between\\ Security and Sustainability}}

\newcommand\copyrighttext{
  \footnotesize \textcopyright Copyright is held by the author/owner. Permission to make digital or hard copies of all or part of this work for personal or classroom use is granted without fee. \\
  This version of this paper was submitted to the Symposium On Usable Privacy and Security (SOUPS)}
\newcommand\copyrightnotice{
  \begin{tikzpicture}[remember picture,overlay]
    \node[anchor=south,yshift=10pt] at (current page.south)
    {\fbox{\parbox{\dimexpr\textwidth-\fboxsep-\fboxrule\relax}{\copyrighttext}}};
  \end{tikzpicture}
}


\def\plainauthor{Keleher, Barrera, and Chiasson}

\author{
{\rm Maxwell Keleher}\\
Carleton University, Canada\\
maxwellkeleher@cmail.carleton.ca
\and
{\rm David Barrera}\\
Carleton University, Canada\\
david.barrera@carleton.ca
\and
{\rm Sonia Chiasson}\\
Carleton University, Canada\\
sonia.chiasson@carleton.ca
} 

\maketitle
\copyrightnotice

\begin{abstract}
    Every year, millions of functional systems become e-waste because users are pressured to send their systems to landfills due to a lack of vendor support and difficulty in recycling. 
    Vendors cite ``cybersecurity'' as the driver for short product support periods, leading to a prevalent, but uninterrogated, belief that cybersecurity and environmental sustainability are fundamentally contradictory; i.e., it is difficult, if not impossible, to build products that are secure, long-lasting, and reusable.
    To understand the nuanced relationship between security and sustainability, we systematically analyze \numpapers papers and distill 155 sustainability guidelines into 12 sustainability themes. These themes enable us to compare the sustainable HCI and sustainable software engineering guidance with that of cybersecurity, identifying points of alignment and tension. 
    We find little evidence of a fundamental tension between these two domains; the few instances of tension can be mitigated through thoughtful consideration of security and sustainability objectives. We also find that sustainability, like usable security, struggles with the myth of users as the weakest link and the individualization of responsibility. Building on these parallels, we argue that the usable security community is well-positioned to integrate sustainability considerations, as both fields share challenges in shifting responsibility from individuals to systemic design. 
\end{abstract}

\section{Introduction}
    According to the 2024 United Nations Global E-waste Monitor~\cite{balde2024global}, e-waste generation is outpacing efforts to collect and recycle it by approximately a factor of five. 
    Vendors' poor support of systems artificially limits their lifespan and significantly contributes to e-waste production~\cite{modarress_fathi_threats_2022}. While there are many reasons for the premature deprecation of functional systems, cybersecurity appears to be a strong motivator (or sometimes culprit) to force users to upgrade to newer systems. What is less clear is whether it is even possible to build systems that are secure, long-lasting, and allow users to reuse and repurpose them for as long as the hardware is functional.
    
    Recent real-world examples suggest that even the largest technology companies are struggling, or unwilling, to consider the environmental impact of their cybersecurity decisions. In March 2025, Google rendered thousands of otherwise functional Chromecasts unusable by letting a security certificate expire~\cite{ChromecastUsersReport2025}. 
    Microsoft's Windows 11 hardware requirements have left millions of PCs unable to receive security updates since support for Windows 10 ended in October 2025~\cite{brankWhatTPMWhy2025}. Most of the abandoned Windows 10 systems are still functional, but users are pressured into buying new computers or risk being exposed to cyberattacks. 
    Android devices typically receive only 1--4 years of security updates\footnote{Companies behind flagship devices such as Google Pixel~8/9/10, Samsung Galaxy~S24/S25, and Fairphone have recently committed to providing 7 years of security updates~\cite{pixel_updates, samsung_updates, fairphone_updates}, but these still make up only a small portion of the global smartphone market.}, after which users must choose between using an unsupported device (i.e., if a vulnerability is found, there may be no official fix to address the vulnerability) or discarding functional hardware.
    Policy and economic pressures may compel companies to act, but the technical challenge of designing systems that are both secure and long-lasting remains fundamentally a human factors problem. 
    It demands the same attention to design that the usable security community has long advocated for when building secure and usable systems. 
    
    We find that the current academic discussion of sustainability and security mirrors early arguments placing security and usability in opposition. \citet{kocksch2023investigating} claim that it is not possible for sustainability and security to be addressed simultaneously. Essentially, ``[sustainability or security] being in focus moves the other to the background''~\cite{kocksch2023investigating}. Other work connecting sustainability and security has uncritically adopted this assumption~\cite{keleherBalancing2025, bradleyEscaping2023}, implying a balance point exists where more security can lead to worse sustainability, and vice-versa, but that both domains cannot be simultaneously prioritized. 
    The sustainable security model presented in the literature to date necessitates separate consideration of both sustainability and security. When viewed as separate concerns, companies are enabled to use security to justify the disregard of sustainability (as seen with the Google, Microsoft, and Android examples).
    
    This paper interrogates whether a fundamental incompatibility exists between sustainability and cybersecurity, as was once assumed to exist for usability and cybersecurity. If such an incompatibility cannot be identified, then what does the relationship between these two domains actually look like?

    \paragraph{Contributions.}
        Through a systematic literature review of guidance for digital sustainability, we extracted 155 guidelines that we distilled into 12 sustainability themes. These themes span the software development life cycle and touch on the variety of stakeholders who impact or are impacted throughout the life cycle. We compare and contrast our 12 sustainability themes with a set of 22 security design principles to examine the idea that there is a fundamental incompatibility between sustainability and security. 
        
        Perhaps unintuitively, we find that cybersecurity and sustainability share many design objectives and seem to be more in alignment than in tension.  
        Our key observations are that: (1) sustainability and security design principles focus on many overlapping objectives and strategies for achieving them, (2) the likely points of tension stem from incomplete, incorrect, or overzealous applications of security principles, and (3) both sustainability and security need to navigate the myth of users being the weakest link.
        In Section~\ref{sec.examples}, we review three real-world examples to demonstrate how our 12 sustainability themes can be used in tandem with security principles to describe security-related sustainability problems in system design and development. 
        Finally, we present a research agenda in Section~\ref{sec:agenda} that describes future work opportunities for usable security to engage with sustainability. 

\section{Background}
   We begin this section by providing our definitions of key terms used in this paper. 
   
   \noindent\textbf{Sustainability:} minimizing the environmental harm caused by e-waste and waste of resources. For example, digital sustainability might look like reusing old computing systems rather than discarding them as e-waste.
   
   \noindent\textbf{Guideline:} referring to design principles, heuristics, guiding questions, or non-functional requirements that capture best practices for system design and development.
   
   \noindent\textbf{System:} broadly referring to digital devices or other computing systems comprised of hardware and/or software.

   \noindent\textbf{End of Life (EoL):} the state when a system is no longer functional, and it is no longer feasible to repair it; however, components of the system might be reused or recycled. 

    \subsection{Computer Security Design Principles}\label{sec:securityprinciples}
        Like sustainability, security is an evolving and ongoing process rather than an objectively verifiable end-state. For this reason, security practitioners must continually evaluate systems (ideally using heuristics and principles, and established best practices), evolving and adapting tools and techniques as adversaries themselves evolve. Ostensibly, security practitioners have developed an intuitive understanding of security best practices, and there exist many sets of principles and heuristics for security (e.g., \cite{cisa2023secure}, \cite{nist2024csf}, \cite{owasp2025top10}). 
        
        No single set of principles will apply universally, but for the purposes of this paper, we have selected the 22 computer security design principles from Chapter 1 of van Oorschot's \textit{Computer Security and the Internet: Tools and Jewels from Malware to Bitcoin}~\cite{van2021computer} because of the breadth of coverage they offer and because of their straightforward and concrete descriptions.
        In contrast, many other principles in the literature were either too high-level or too task-specific for our purposes. We believe that van Oorschot's security principles appropriately balance specificity and abstraction. They are sufficiently broad to cover most relevant aspects of computer security (i.e., they apply to systems security, network security, usable security, etc.). At the same time, each principle is appropriately narrow to avoid redundancy and ambiguity. These security principles are suited for the variety of development and design contexts we encountered in our review of sustainable design principles. We provide a summary of each principle in Table~\ref{tab:secprinciples}, and the full definitions in Appendix~\ref{app:secprinciples}. 

    \begingroup
        \renewcommand*{\arraystretch}{1.2}
                \begin{table*}[tb]
                \footnotesize
                \rowcolors{1}{white}{gray!20}
                \centering
                    \begin{tabular}{p{1.9in}p{4.75in}}
                    \toprule
                   \textbf{Security Principle} & \textbf{Brief Description} \\ 
                   \hline
                  \securitybydesign & Build security in from the start of the development cycle \\
                  
                  \designforevolution & Design base architectures, mechanisms, and protocols to support evolution, and regularly reevaluate security mechanisms in light of evolving threats \\ \hline
        
                  \simplicityandnecessity & Keep designs as simple and as small as possible. \\
                  \safedefaults & Use safe default settings, including fail-safe designs. \\
                  \opendesign & Invite and encourage open review and analysis. \\
                  \completemediation & Verify proper authority for each access to every object. \\
                  \isolatedcompartments & Compartmentalize system components using strong isolation. \\
                  \leastprivilege & Allocate fewest privileges necessary and for shortest duration necessary. \\
                  \modulardesign & Favor object-oriented designs that segregate privileges. \\
                  \smalltrustedbases & Strive for small code sizes in components that must be trusted. \\
                  \timetestedtools & Rely on time-tested, expert-built security tools and protocols. \\
                  \leastsurprise & Design security mechanisms and user interfaces that behave as users expect. \\
                  \userbuyin & Design security mechanisms where users are motivated to follow the safest, easiest path. \\
                  \sufficientworkfactor & Tune security mechanisms so that the cost to defeat it clearly exceeds anticipated resources of adversaries. \\
                  \defenseindepth & Build defences in multiple layers backing each other up. \\
                  \evidenceproduction & Record system activities through event logs and monitoring tools. \\
                  \datatypevalidation & Verify that all received data meets expected properties. \\
                  \remnantremoval & On termination of a session, remove all traces of sensitive data. \\
                  \trustanchorjustification & Ensure confidence in any base point of assumed trust and verify trust assumptions. \\
                  \idependentconfirmation & Use simple, independent cross-checks to increase confidence in code or data. \\
                  \requestresponseintegrity & Verify that responses match requests in distributed protocols. \\
                  \reluctantallocation & Be reluctant to allocate resources with unauthenticated external agents. \\
         
        \bottomrule
                    \end{tabular}
                    \caption{Summary of van Oorschot's security principles~\cite{van2021computer}. HP\# indicates a high-level principle, P\# indicates a principle.  Full descriptions are available in Appendix~\ref{app:secprinciples}.}
                \label{tab:secprinciples}
                \end{table*}
        \endgroup

    \subsection{Related Work}
        To the best of our knowledge, our review covers the broadest scope of digital sustainability guidelines. 
        Several previous sustainability literature reviews have focused on HCI and design~\cite{disalvoMappingLandscapeSustainable2010, dealmeidanerisSystematicReviewSustainability2014, sierra-fontalvoDeepDiveAddressing2023, moustafaExaminingSustainableProduct2025, bremerHaveWeTaken2022a, hanssonDecadeSustainableHCI2021a, poderiParticipatoryDesignSustainability2018}. 
        Additionally, many previous literature reviews have investigated connections between sustainability and software engineering~\cite{garcia-mirelesInteractionsEnvironmentalSustainability2018, wolframSustainabilitySoftwareEngineering2017, penzenstadlerSystematicMappingStudy2014,berntsenSustainabilitySoftwareEngineering2017, garcia-mirelesEnvironmentalSustainabilitySoftware2017, marimuthuSoftwareEngineeringAspects2017, anwarGreenerSoftwareEngineering2017, moisesdesouzaSocialSustainabilityApproaches2024, raibuletSelfadaptationSustainableSoftware2024, dawoodMappingAnalysisOpen2019, haiderFactorsInfluencingSustainability2024, swachaModelsSustainableSoftware2022}. 
        Previous reviews of literature related to digital sustainability have focused on leveraging technology as a solution to environmental sustainability issues~\cite{abubakarIoTUN2025, giudiciPersuasive2025, chatterjeeComputational2020}.
        One of the past literature reviews interrogates the sustainability of technology and specifically focuses on the carbon emission or energy efficiency of technology~\cite{danushiCarbonEfficient2025}. 
        We acknowledge that minimizing emissions and energy consumption are important goals; however, we seek to obtain a more holistic understanding of designing for sustainability. 

    \subsection{Sustainability and Security}
        There is an emerging interest in considering the sustainability impacts of security decisions or practices. 
        Although a security vulnerability on its own may not trigger end users to discard their device~\cite{keleherBalancing2025}, security researchers have recently begun exploring ways of extending device lifespans by extending security update support after it is no longer provided by the device manufacturer~\cite{cobbinaEnhancing2026, bradleyEscaping2023}. 

        In a 2023 workshop paper titled \textit{Investigating the Sustainable-Cybersecurity Nexus in HCI as a Practical Problem}~\cite{kocksch2023investigating}, \citeauthor{kocksch2023investigating} discuss the tensions that they identified via ethnographic fieldwork of IT systems. They conducted one ethnographic study of a data center at a German university, and another of IT systems at small and medium-sized businesses in Denmark. Based on observations from these two ethnographic studies, \citeauthor{kocksch2023investigating} extrapolate that sustainability is fundamentally in tension with cybersecurity, and they argue that there is a need to address the tensions arising between the two issues. In their studies, they observe that when either sustainability or cybersecurity is the focus, then the other is deprioritized. They suggest it is necessary to alternately address sustainability and security to balance the two objectives. 

    \subsection{Research Direction and Scope} 
        In this paper, we seek to understand where the alleged points of tension between sustainability and security arise. Moreover, we seek to understand where, if at all, there is complementary or synergistic guidance bridging the two fields. To do so, we conduct a new holistic review of sustainability guidelines that includes both the design and development of technology in its scope. In contrast to past reviews, we are interested in work that focuses on making digital technology itself more sustainable, rather than using digital technology to achieve sustainability in other areas. 

\section{Method}
        Our literature review followed a citation-chasing approach~\cite{cooper2019handbook}. We reviewed the references of a given paper (backwards search) and the other papers that cite it (forwards search). We use the term \textbf{references} to indicate papers that are \textit{cited by} a given paper (i.e., the list of references at the end of the paper), and the term \textbf{citations} to refer to papers that \textit{cite} a given paper. 
        
        We started our citation chasing with an initial set of seven seed papers summarized in Table~\ref{tab:seedpapers}. We identified these papers from an initial hand-search~\cite{cooper2019handbook} of the literature. We selected recent reviews of sustainable HCI (SHCI)~\cite{hanssonDecadeSustainableHCI2021a, bremerHaveWeTaken2022a} and sustainable software engineering~\cite{mouraoGreenSustainableSoftware2018a, penzenstadlerSafety2014} domains, and foundational, highly cited contributions~\cite{cohnConvivial2016, blevisSustainableInteractionDesign2007a, mankoffEnvironmental2007}.
        \citet{blevisSustainableInteractionDesign2007a} presents the earliest set of sustainable design guidelines we could find, and, along with \citet{mankoffEnvironmental2007}'s work, appears to mark the origin point of SHCI research. 
        While the other five seed papers did not themselves present sustainability guidelines, they provided entry points to topics or fields of study relevant to sustainable design and engineering (e.g., managing EoL, SHCI, and Green IT).

    \subsection{Paper Collection}
        \begin{figure*}[tb]
            \centering
            \includegraphics[width=.6\linewidth]{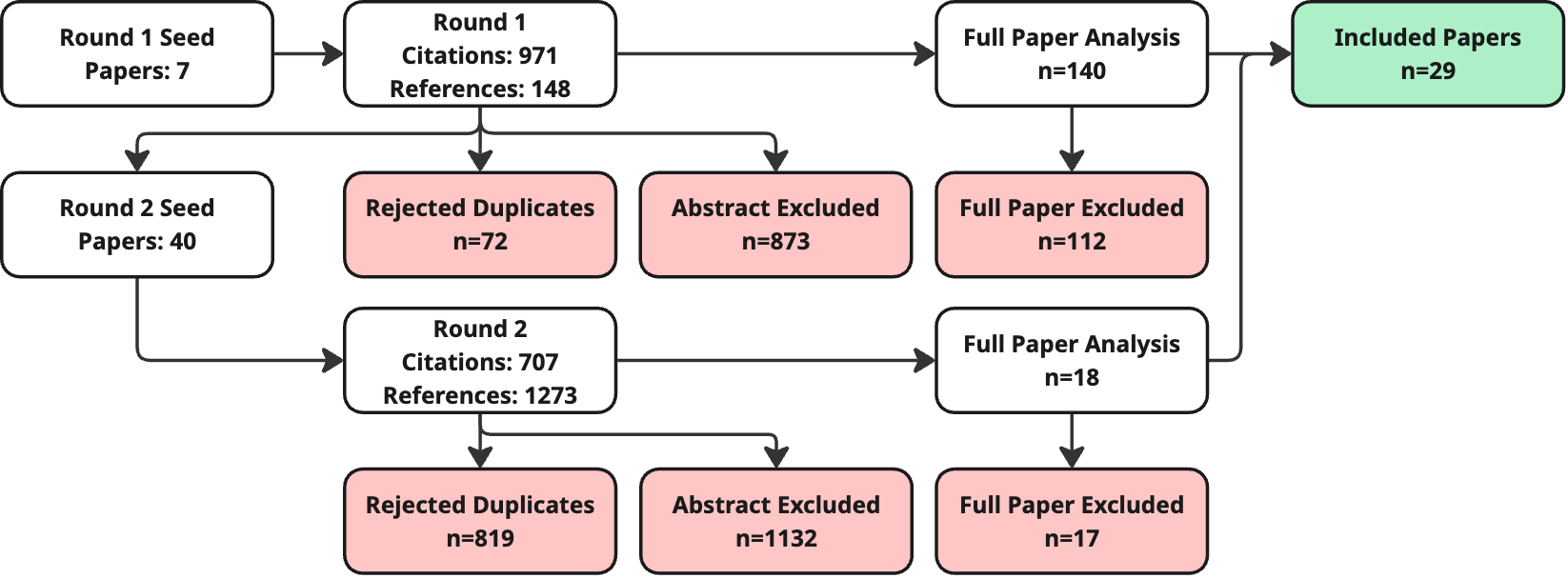}
            \caption{Overview paper collection and filtering process.}
            \label{fig:ReviewProcess}
        \end{figure*}

        Our paper compilation process is summarized in Figure~\ref{fig:ReviewProcess}.
        We used citationchaser~\cite{haddawayCitationchaserPackageForward2021}, a tool which compiles the references and citations for a set of papers, to construct our list of papers to review. 
        We conducted the paper search in October 2025.
        
        \paragraph{Filtering and Exclusion Criteria.}
            When assessing papers from the citation-chasing phase, our inclusion criteria were papers that presented guidelines related to sustainability, repair and maintenance, or EoL. Our exclusion criteria consisted of papers that focused on: energy efficiency or carbon emissions as the exclusive or primary concern, promoting end-user awareness of sustainability or sustainable behavior, or papers focused on machine learning or artificial intelligence. 
        
        \paragraph{Round 1.}
            Round 1 of the forward-backward search from the seed papers yielded 1119 papers (148 references and 971 citations).
            In the initial review, we considered the title and abstract of each paper and filtered papers based on our inclusion and exclusion criteria. 873 papers were rejected during the review of the abstracts, and 72 duplicates were removed\footnote{In Round 1, duplicates occurred because of indexing errors in the online literature databases and because of the different seed papers respectively referencing and being cited by the same paper.}. 
            Afterwards, we reviewed the remaining 140 papers (7 seed papers and the remaining 133 collected papers) to ensure that they presented novel, relevant guidelines. 
            112 were rejected during the review of the full papers because they did not include guidelines, only repeated guidelines from other sources, or included guidelines that were not relevant to our analysis (e.g., guiding education rather than guiding design and development). 
            We ended up with 28 papers on which we conducted a first round of thematic analysis (process detailed in Sec \ref{sec:analysis}). 

        \paragraph{Round 2.}
            During Round 1, we identified additional literature review papers that reviewed guidelines and workshop proceedings that might lead to new guidelines, even though they themselves did not meet our inclusion criteria. 
            We used these literature reviews and workshop proceedings as seed papers for Round 2 of paper collection (n=40). In Round 2, these seed papers yielded 1980 papers (1273 references and 707 citations). We filtered out 819 duplicate (i.e., arising from different seed papers) papers and then filtered out 1132 papers during abstract review. Another 17 were filtered out during the review of the full papers, resulting in the inclusion of one paper for analysis for a total of \numpapers. We incorporated this paper into our thematic analysis.
            We stopped our search after two rounds because Round 2 yielded only one included paper, and we encountered no new codes or themes during analysis.

    \subsection{Analysis}\label{sec:analysis}
        After each round of paper collection, we extracted the text of the guidelines (and the provided descriptions or definitions if available) from the included papers to a spreadsheet as the content for inductive thematic analysis~\cite{braunThematicAnalysisPractical2022}. In total, we extracted 155 guidelines. 
        We open-coded the extracted guidelines, and, after each round of collection, we reviewed and merged the open codes to create themes that describe common patterns in the sustainability guidelines. 
        This analysis yielded 12 sustainability themes (defined in Section~\ref{sec:review}). Our 12 themes are comprised of 123 of the extracted guidelines. 32 guidelines were coded to a miscellaneous group because they were not related to any of the 12 themes, and could not be collected into an additional theme. 
        
        In the second phase of our analysis, we iteratively compared our sustainability themes against van Oorschot's~\cite{van2021computer} security principles (Table~\ref{tab:secprinciples}, Appendix~\ref{app:secprinciples}). 
        We considered each security principle in relation to each sustainability theme, and determined whether the security principle was in alignment or tension with the objectives of the themes. 
        For example, the sustainability theme \usability (see Section~\ref{sec:usage}) most obviously aligns with the security principles \leastsurprise and \userbuyin, matching users' mental models of system security. Avoiding confusing system behaviors is particularly important as it may push users to dispose of a system with a fixable vulnerability. We also found that the security principle, \simplicityandnecessity aligns with this theme because a simple system is easier for a user to understand and learn.
        Finally, we noticed that the security principle \reluctantallocation can be in tension with the objective of usability because prompting users to provide permission for connections and processes might frustrate them.

        The first author conducted the majority of the analysis, with frequent in-depth discussions with the other two authors, leading to several iterations refining the results and insights. Iterations concluded once all authors agreed on the themes as well as the points of alignment and tension. 
        In Section~\ref{sec:review}, we describe each sustainability theme and its relationships with the security design guidelines. Note that cells marked with alignment or tension in Tables~\ref{tab:designdev}, \ref{tab:usage}, \ref{tab:disposal} can have varying degrees of significance; the relative number of marked cells does not imply an objectively more significant relationship. 
    \paragraph{Researcher Backgrounds.}
        The research team recognizes that qualitative methodologies are inherently subjective and that the unique perspective of the researchers doing the analysis contributes to the value of the work. 
        The majority of the analysis was performed by the first author, Maxwell Keleher. He is a computer science PhD candidate with a focus on human-computer interaction and has professional experience as a software developer and as a human-factors researcher. He has significant experience with qualitative analysis from both his academic and professional backgrounds. 
        David: Dr. David Barrera is an associate professor of computer science with a research background in computer and network security, as well as security and environmental sustainability.
        Sonia: Dr. Sonia Chiasson is a Professor of computer science with a research background in usable security and privacy with quantitative and qualitative.
        
\section{Literature Review and Analysis}\label{sec:review}
    In this section, we define the 12 sustainability themes (Table~\ref{tab:sustainabilityThemes}) that we developed through thematic analysis of 155 sustainability guidelines from \numpapers papers. These sustainability themes correspond to three phases of the software development life cycle (1: design and development, 2: usage, and 3: disposal and end of life). For each phase, we describe the points of alignment and tension between our sustainability themes, and van Oorschot's security principles~\cite{van2021computer}, summarized in Table~\ref{tab:secprinciples}.

    \begin{table*}[h]
        \centering
        \footnotesize
        
        \begin{tabular}{lllc}
        \toprule
            \textbf{Life Cycle Phase} &\textbf{ Sustainability Themes} & \textbf{Included Papers} & \textbf{\# Guidelines}\\ \midrule
            Design and Development (Sec \ref{sec:designanddev}) & \simplifyandreduce & \cite{blevisAdvancingSustainableInteraction2006,hakanssonBeingGreenSimple2013,remyBridgingTheoryPracticeGap2015,andersonCommunicatingSustainability2008,gegenbauerInspiringDesignLongerlived2012,englhardtIncorporatingSustainabilityElectronics2025,hazasSustainabilityDoesNot2012,remyLimitsEvaluatingSustainability2017,jung2010conceptualizations} & 10\\
            & \durabilityandlongevity & \cite{wakkarySustainableIdentityCreativity2009,chenStrategyLimitsawareComputing2016,blevisAdvancingSustainableInteraction2006,jainChallengesEnvironmentalDesign2002,khalifehExploringNexusSustainability2023,blevisFurtherConnectingSustainable2017,khalifehIncorporatingSustainabilitySoftware2020,beckerSustainabilityDesignSoftware2015} & 9\\
            & \collaboration & \cite{blevisAdvancingSustainableInteraction2006,tomlinsonCollapseInformaticsAugmenting2012,jung2010conceptualizations,nystromExploringSustainableHCI2018,khalifehExploringNexusSustainability2023,huhFindingLostTreasure2010,blevisFurtherConnectingSustainable2017,khalifehIncorporatingSustainabilitySoftware2020,odomPersonalInventoriesDurable2008,phamShapeREMultiDimensionalRepresentation2020,beckerSustainabilityDesignSoftware2015,remyLimitsEvaluatingSustainability2017,preistUnderstandingMitigatingEffects2016} & 16\\
            & \adaptandaugment & \cite{remyBridgingTheoryPracticeGap2015,jainChallengesEnvironmentalDesign2002,tomlinsonCollapseInformaticsAugmenting2012,jung2010conceptualizations,nystromExploringSustainableHCI2018,odomPersonalInventoriesDurable2008,luUnmakingElectronicWaste2024} & 9\\
            & \efficiency & \cite{jainChallengesEnvironmentalDesign2002,andersonCommunicatingSustainability2008,khalifehExploringNexusSustainability2023,khalifehIncorporatingSustainabilitySoftware2020,phamShapeREMultiDimensionalRepresentation2020,preistUnderstandingMitigatingEffects2016} & 13\\
            & \context & \cite{chenStrategyLimitsawareComputing2016,penzenstadlerToolkitSESustainability2015,hakanssonBeingGreenSimple2013,andersonCommunicatingSustainability2008,blevisFurtherConnectingSustainable2017,beckerSustainabilityDesignSoftware2015,remyLimitsEvaluatingSustainability2017,burgerCapabilityApproachSustainability2011} & 19\\ \midrule
            
            Usage (Sec \ref{sec:usage})& \sususerbuyin & \cite{blevisAdvancingSustainableInteraction2006,remyBridgingTheoryPracticeGap2015,jung2010conceptualizations,gegenbauerInspiringDesignLongerlived2012,blevisSustainableInteractionDesign2007a} & 10 \\
            & \usability & \cite{hakanssonBeingGreenSimple2013,tomlinsonCollapseInformaticsAugmenting2012,nystromExploringSustainableHCI2018,khalifehExploringNexusSustainability2023,niemelaHDD20214,khalifehIncorporatingSustainabilitySoftware2020,phamShapeREMultiDimensionalRepresentation2020} & 9 \\ 
            & \repairandmaintain & \cite{blevisAdvancingSustainableInteraction2006,tomlinsonCollapseInformaticsAugmenting2012,khalifehExploringNexusSustainability2023,khalifehIncorporatingSustainabilitySoftware2020,phamShapeREMultiDimensionalRepresentation2020,chengTransientInternetThings2023,luUnmakingElectronicWaste2024,kilicUserCentredRepairCurrent2024} & 13 \\\midrule
                        
            Disposal and End of Life (Sec \ref{sec:disposal}) & \reuse & \cite{blevisAdvancingSustainableInteraction2006,tomlinsonCollapseInformaticsAugmenting2012,pierceSecondhandInteractionsInvestigating2011,blevisSustainableInteractionDesign2007a,chengTransientInternetThings2023} & 10 \\
             & \recycle & \cite{blevisAdvancingSustainableInteraction2006,chengTransientInternetThings2023} & 2\\
             & \degrade & \cite{blevisAdvancingSustainableInteraction2006,jainChallengesEnvironmentalDesign2002,chengTransientInternetThings2023} &  3\\ \bottomrule

        \end{tabular}
        \caption{The sustainable themes that emerged from the extracted guidelines and the associated papers. The ``\# Guidelines'' column indicates the number of guidelines that were coded into each theme.}
        \label{tab:sustainabilityThemes}
    \end{table*}

    \subsection{Design and Development} \label{sec:designanddev}

        \begingroup
        \renewcommand*{\arraystretch}{1.2}
                \begin{table*}[]
                \scriptsize
                \centering
                    \begin{tabular}{p{1.25in}|lllllllllllll|llll|l|}
        
                        \multicolumn{1}{c|}{} & P1 & P2 & P3 & P5 & P6 & P7 & P8 & P9 & P10 & P11 & P13 & P16 & P18 & P4 & P12 & P14 & P20 & HP2 \\ 
                        \hline
                        
                        \simplifyandreduce & \cellAligns{+} & ~ & ~ & ~ & \cellAligns{+} & \cellAligns{+} & \cellAligns{+} & ~ & ~ & ~ & ~ & \cellAligns{+} & ~ & ~ & \cellConflicts{-} & \cellConflicts{-} & \cellAligns{+} & ~ \\ 
                        \durabilityandlongevity & \cellAligns{+} & \cellAligns{+} & ~ & ~ & ~ & ~ & ~ & \cellAligns{+} & ~ & ~ & \cellAligns{+} & ~ & ~ & \cellConflicts{-} & ~ & ~ & ~ & \cellAligns{+} \\ 
                        \collaboration & ~ & ~ & \cellAligns{+} & ~ & ~ & \cellAligns{+} & ~ & \cellAligns{+} & ~ & ~ & ~ & ~ & \cellAligns{+} & \cellConflicts{-} & ~ & ~ & \cellConflicts{-} & ~ \\ 
                        \adaptandaugment & ~ & ~ & \cellAligns{+} & \cellAligns{+} & ~ & \cellAligns{+} & ~ & ~ & ~ & ~ & ~ & ~ & ~ & ~ & ~ & ~ & \cellConflicts{-} & \cellAligns{+} \\ 
                        \efficiency & \cellAligns{+} & ~ & ~ & ~ & ~ & ~ & ~ & ~ & ~ & ~ & ~ & ~ & ~ & ~ & \cellConflicts{-} & ~ & ~ & ~ \\ 
                        \context & ~ & ~ & ~ & ~ & ~ & ~ & ~ & ~ & \cellAligns{+} & \cellAligns{+} & ~ & ~ & ~ & ~ & ~ & ~ & ~ & ~ \\
        
                        \hline
                        \multicolumn{3}{l}{\simplicityandnecessity}
                        & \multicolumn{6}{l} {\smalltrustedbases}       
                        & \multicolumn{6}{l}{\idependentconfirmation} \\
                        
                        \multicolumn{3}{l}{\safedefaults}          
                        & \multicolumn{6}{l} {\timetestedtools}   
                        & \multicolumn{6}{l}{\completemediation}  \\
                        
                        \multicolumn{3}{l}{\opendesign}            
                        & \multicolumn{6}{l} {\leastsurprise}     
                        & \multicolumn{6}{l}{\sufficientworkfactor} \\
                        
                        \multicolumn{3}{l}{\isolatedcompartments}        
                        & \multicolumn{6}{l} {\userbuyin}       
                        & \multicolumn{6}{l}{\evidenceproduction} \\
                        
                        \multicolumn{3}{l}{\leastprivilege}        
                        & \multicolumn{6}{l} {\defenseindepth}       
                        & \multicolumn{6}{l}{\reluctantallocation} \\

                        \multicolumn{3}{l}{\modulardesign}        
                        & \multicolumn{6}{l} {\remnantremoval}       
                        & \multicolumn{6}{l}{\designforevolution} \\
            
                    \end{tabular}
                    \caption{Matrix showing relationships between sustainability themes from the review and van Oorschot's security principles~\cite{van2021computer} for the Design and Development phase. Green ``+'' cells indicate positive relationships between the security and sustainability guidelines. Red ``-'' cells indicate negative relationships or tension between the security and sustainability guidelines.}
                \label{tab:designdev}
                \end{table*}
        \endgroup

        Over half of the sustainability guidelines identified in our review pertained to practices intended for the design and development phase of the product life cycle. 
        Similar to the importance of considering security throughout the design processes, there is value in considering sustainability from the beginning. The decisions made in the Design and Development phase can have a significant impact on the sustainability and security of a system. 
        For example, a system designed in a complex way may be more confusing for users and harder for them to maintain, which has negative implications for both security and sustainability; a complex system might also be more difficult for users to recycle in a way that maintains security and privacy. 
        
        Next, we define each of the six sustainability themes relating to the Design and Development phase, then discuss our analysis of how these themes relate to van Oorschot's security design principles~\cite{van2021computer} (represented in Table~\ref{tab:designdev}). 
        
        \paragraph{\simplifyandreduce:}
            This theme emphasizes managing constraints on available resources or otherwise minimizing the resources necessary to build and operate a system. Designing systems to reduce complexity and minimize resource usage makes them easier to maintain for long lifespans and reduces e-waste once users eventually dispose of them. 

        \paragraph{\durabilityandlongevity:}
            This theme is about designing systems in ways that maximize their lifespan. For example, it encourages designing systems to be resilient and reliable because these systems are more likely to remain in use for a long time. By considering the expected lifespan of the systems they are making, designers and developers can mitigate the likelihood that the system will be disposed of prematurely. 
            
        \paragraph{\collaboration:}
            This theme emphasizes following open design practices or leveraging shared resources and infrastructure. Systems that follow open design are easier to repair and maintain, resulting in longer lifespans. Moreover, they are more likely to be able to switch to third-party infrastructure should the original manufacturer end support. Systems that leverage shared infrastructure or resources eliminate duplicate effort and also drive infrastructure reliability. 
            
        \paragraph{\adaptandaugment:}
            This theme relates to designing systems that can adapt to changing conditions or can be augmented after initial deployment. Adaptable and augmentable systems are likely to have longer lifespans and minimize the risk of being discarded if users' needs change. 
        
        \paragraph{\efficiency:}
            This theme relates to reducing waste in the design and development process and designing systems to be more efficient themselves. It includes designing systems to make use of digital and physical resources more efficiently, minimizing e-waste and energy expenditure. 

        \paragraph{\context:}
            To properly address sustainability issues, there must be appropriate context gathering during the design and development of a system. By considering a greater variety of stakeholders (e.g, people who are not users but may experience consequences from e-waste), and tailoring the design to the context in which the system is expected to be used, systems are less likely to be wasteful. 

        \subsubsection{Alignment in the Design and Development Phase}
            We found that both sustainability and cybersecurity advocate for, and benefit from, a set of shared best practices: embracing simple, compartmentalized designs, following open design practices, and embracing adaptability. 
            Both domains caution against monolithic systems because they are harder to maintain and risk entire systems being compromised by the failure of a single component. Simple, modular systems are less wasteful, easier to secure, and allow for extending a system's lifespan by replacing individual components rather than discarding the whole. For example, security principles like \modulardesign and \isolatedcompartments compartmentalize system design to achieve simplicity (aligns with the \simplifyandreduce theme) and compartmentalized designs are easier to modify in the future (alignment with the \adaptandaugment theme).

            \begin{tcolorbox} [width=\linewidth, sharp corners=all, colback=white!90!green]
                \paragraph{Key Alignment:} Simple, modular systems are easier for security experts to maintain \textit{and} are more likely to have long life spans. 
            \end{tcolorbox}

            From a security perspective, open design enables external review that can identify overlooked vulnerabilities. For sustainability, open systems encourage collaborative reuse of reliable components and are easier to repair. Both benefit from adopting time-tested, publicly available tools that are well-understood and less likely to limit system lifespan.
            
            \begin{tcolorbox} [width=\linewidth, sharp corners=all, colback=white!90!green]
                \paragraph{Key Alignment:} Open design encourages better system sustainability \textit{and} security. 
            \end{tcolorbox}

        \subsubsection{Tension in the Design and Development Phase}
            Through our analysis, we identified that a narrow interpretation of certain security principles can tension with the objectives of our sustainability themes. 
            For example, the security principle \sufficientworkfactor is in tension with the sustainability theme of \efficiency because it achieves security by requiring more computational effort than an attacker is anticipated to possess. In security, it is sometimes necessary to leverage inefficiency to discourage malicious interactions with systems. When assessing the sustainability of systems with these features, it is important to incorporate the likelihood of inefficiency occurring in real-world use. Similarly, \evidenceproduction is wasteful because it involves generating and storing extra data (to be accessed in the event of a compromise that may or may not take place). For \sufficientworkfactor and \evidenceproduction, we believe it is possible to negotiate a compromise between security and sustainability objectives, and that there should be sufficient justification provided for security decisions that result in inefficiency or wastefulness. 

            \begin{tcolorbox} [width=\linewidth, sharp corners=all, colback=white!90!red]
                \paragraph{Key Tension:} Infrequent inefficiency can be necessary for preventing system misuse. 
            \end{tcolorbox}

    \subsection{Usage} \label{sec:usage}
        \begingroup
        \renewcommand*{\arraystretch}{1.2}
                \begin{table}[]
                \scriptsize
                \centering
                    \begin{tabular}{p{.8in}|lllllll|l|}
        
                        \multicolumn{1}{c|}{} & P1 & P3 & P5 & P9 & P10 & P11 & P20 & HP2 \\ 
                        \hline
                        
                        \sususerbuyin & ~ & ~ & ~ & ~ & \cellAligns{+} & \cellAligns{+} & ~ & ~ \\
                        \usability & \cellAligns{+} & ~ & ~ & ~ & \cellAligns{+} & \cellAligns{+} & \cellConflicts{-} & ~ \\ 
                        \repairandmaintain & ~ & \cellAligns{+} & \cellAligns{+} & \cellAligns{+} & ~ & ~ & \cellConflicts{-} & \cellAligns{+} \\ 
                        \hline
            
            \multicolumn{4}{l}{\simplicityandnecessity} & \multicolumn{5}{l} {\leastsurprise} \\  
            \multicolumn{4}{l}  {\opendesign} & \multicolumn{5}{l}  {\userbuyin}\\
            \multicolumn{4}{l}  {\isolatedcompartments} & \multicolumn{5}{l} {\reluctantallocation} \\
            \multicolumn{4}{l} {\timetestedtools} & \multicolumn{5}{l} {\designforevolution} \\
            
                    \end{tabular}
                    \caption{Matrix showing relationships between sustainability themes from the review and van Oorschot's security principles~\cite{van2021computer} for the Usage phase. Green ``+'' cells indicate positive relationships between the security and sustainability guidelines. Red ``-'' cells indicate negative relationships or tension between the security and sustainability guidelines. }
                \label{tab:usage}
                \end{table}
        \endgroup
    
       The usage phase of a system's life cycle focuses on the end-user themes and practices. The sustainability guidelines associated with these themes tend to offer guidance on how designers and developers can build systems that align with end users' expectations and promote sustainable behaviors within system ownership communities. 
       
       Indeed, there is a similar need in cybersecurity to align design decisions with end users' expectations to achieve desired security behaviors and outcomes. Security is typically not a primary goal for end users~\cite{Sasse2005UsableSW}, and the same seems true for sustainability. Thus, designers and developers must carefully consider how to encourage users to adopt behaviors they are not intrinsically motivated to pursue. 

       We define three sustainability themes related to the usage phase that emerged from our analysis, and then describe alignment and tension between these themes and the security principles by \citet{van2021computer} (represented in Table~\ref{tab:usage}). 

        \paragraph{\sususerbuyin:}
            This theme focuses on how to engage the user in sustainable practices and how to encourage long-term use of their existing systems. A key insight across the guidelines is that user behavior and perceptions have a significant impact on the sustainability of the system. 
            In other words, users will use systems for longer and maintain them if they perceive them to be durable or valuable. Conversely, users do not care about systems that they perceive to be cheap and disposable.

        \paragraph{\usability:}
            This theme embraces the HCI ideals relating to usability, including promoting user-centered design methodologies and involving users in co-design. Systems designed with these approaches are more likely to be intuitive and efficient to use, and therefore, they are less likely to be prematurely discarded. 

        \paragraph{\repairandmaintain:}
            This theme encourages features that extend system lifespan, such as the capacity to easily repair and maintain systems. 
            Systems following these guidelines are easy to diagnose and have reputable repair guides. Any components that impact lifespan are easy to repair or replace. When designed in this way, systems have a longer lifespan.

        \subsubsection{Alignment in the Usage Phase}
            Through our analysis, we identified a commonality between sustainability and cybersecurity guidelines that mutually focus on the importance of user-centered approaches.
            We found that in both domains, user buy-in is necessary because end-user behaviors can either significantly contribute to or detract from achieving the overarching sustainability or security goal. 
            In both cases, predictability is key to encouraging desirable behaviors from users (i.e., security or sustainability practices). 
            Moreover, efforts to improve the learnability of systems could provide both security and sustainability benefits. 

            Several security principles can contribute to more sustainable designs (see cells labelled with ``+'' in Table~\ref{tab:usage}). For example, \opendesign and \isolatedcompartments can lead to systems that are easier to diagnose and repair. \timetestedtools can help create better documentation and increase the availability of components.
            Additionally, \simplicityandnecessity can reduce complexity and help increase longevity.
            Finally, usage friction is reduced by following \userbuyin and \leastsurprise, which encourage users to keep using their systems, thus extending longevity.
            
            \begin{tcolorbox} [width=\linewidth, sharp corners=all, colback=white!90!green]
            \paragraph{Key Alignment:} Usability is key to encouraging secure \textit{and} sustainable user behaviors.
            \end{tcolorbox}

        \subsubsection{Tension in the Usage Phase}
            In the usage phase, \reluctantallocation was the only security principle in tension (see ``-'' cells in Table~\ref{tab:usage}). Strict adherence to this principle may introduce extra friction for users by requiring them to constantly permit the sharing of resources or permissions, and push users towards discarding a system because they find it frustrating to use.

            The ability to repair systems is a critical consideration to address sustainability, but it can present unintended security risks. The integration of compromised components or modules during repair could lead to issues and to system disposal. For example, replacing a component with one that has outdated firmware might introduce vulnerabilities, break the system, and lead to premature disposal. Alternatively, someone might install malicious hardware to surveil the owner of a system. There are also documented instances of repair technicians snooping on systems that customers brought for repair~\cite{kampfWeCaughtTechnicians2023}, which may make users leery of choosing to repair.
            Similarly, easily repaired or modified systems could open up opportunities for technology-facilitated abuse\footnote{Technology-facilitated abuse (TFA) is a type of intimate partner violence where the abuser stalks or harasses with the assistance of digital technology (e.g., doorbell cameras, smart trackers) or illicitly accesses their partner's accounts or devices~\cite{rogersTechnology-Facilitated2023}.}.
            Verifying the legitimacy of repairs to systems (whether done by yourself or others) is crucial to ensure security. Sustainability and security experts must collaborate to ensure repairability does not introduce unnecessary security risks.
            \begin{tcolorbox} [width=\linewidth, sharp corners=all, colback=white!90!red]
            \paragraph{Key Tension:} Bad actors can exploit usable repair. 
            \end{tcolorbox}
    
    \subsection{Disposal and End of Life} \label{sec:disposal}
        \begingroup
        \renewcommand*{\arraystretch}{1.2}
                \begin{table}[]
                \scriptsize
                \centering
                    \begin{tabular}{p{.8in}|lllllll|l|}
        
                        \multicolumn{1}{c|}{} & P1 & P2 & P5 & P7 & P16 & P14 & P20 & HP2 \\ 
                        \hline
                        
                        \reuse & ~ & ~ & \cellAligns{+} & \cellAligns{+} & \cellAligns{+} & \cellConflicts{-} & \cellConflicts{-} & ~ \\ 
                        \recycle & \cellAligns{+} & ~ & ~ & ~ & \cellAligns{+} & \cellConflicts{-} & ~ & ~ \\ 
                        \degrade & ~ & \cellAligns{+} & ~ & ~ & \cellAligns{+} & \cellConflicts{-} & ~ & \cellAligns{+} \\ 
        
        \hline
            \multicolumn{4}{l}{\simplicityandnecessity} & \multicolumn{5}{l} {\remnantremoval} \\  
            \multicolumn{4}{l}  {\safedefaults} & \multicolumn{5}{l}  {\evidenceproduction}\\
            \multicolumn{4}{l}  {\isolatedcompartments} & \multicolumn{5}{l} {\reluctantallocation} \\
            \multicolumn{4}{l} {\modulardesign} & \multicolumn{5}{l} {\designforevolution} \\
            
                    \end{tabular}
                    \caption{Matrix showing relationships between sustainability themes from the review and van Oorschot's security principles~\cite{van2021computer} for the Disposal and End of Life phase. Green ``+'' cells indicate positive relationships between the security and sustainability guidelines. Red ``-'' cells indicate negative relationships or tension between the security and sustainability guidelines.}
                \label{tab:disposal}
                \end{table}
        \endgroup

        Degradation is an inevitable aspect of a digital system's life cycle. While the owners of the system typically decide how to dispose of it (e.g., reuse, recycle, send to landfill), the ease with which they can make sustainable choices is often determined by upstream decisions by designers or developers. Moreover, limitations in the public infrastructure available to owners may also affect end users' ability to make sustainable choices, so public service providers and those who manage municipal infrastructure may also be relevant stakeholders.
        We define three sustainability themes related to the disposal and end of life phase that emerged from our analysis, and then describe the alignment and tension between these themes and the security principles by \citet{van2021computer} (represented in Table~\ref{tab:disposal}). 
        
        \paragraph{\reuse:}
            This theme relates to the reuse of systems after an initial phase of usage. Systems that reuse old components, or that are designed so that their components can be easily reused, divert still usable components from becoming e-waste. Moreover, systems that are designed to facilitate easy ownership transfer can have long lifespans even if they stop being used by the original owner. 

        \paragraph{\recycle:}
            Recycling involves breaking down a system and processing its components or material into new ones. Systems that are designed to be easily recycled, or that make use of recycled materials in their design, minimize the generation of e-waste. 

        \paragraph{\degrade:}    
            All systems will inevitably reach EoL, but thoughtful design around how a system reaches and behaves at EoL can minimize negative sustainability impacts. Examples of graceful degradation strategies include: self-destroying data~\cite{jainChallengesEnvironmentalDesign2002}, biodegradable components~\cite{chengTransientInternetThings2023},  staggered deprecation of functionality~\cite{gouldDealingDigitalService2022}, and migration to read-only states~\cite{gouldDealingDigitalService2022}. 
        
        \subsubsection{Alignment in the Disposal and End of Life Phase}
            In our analysis of the security principles and sustainability themes for the disposal phase, we found that certain security principles supported the sustainability objectives of minimizing the amount of e-waste from discarded systems. 
            For example, the sustainability theme \reuse encourages the use of existing components, which is made easier by the security principles \isolatedcompartments and \modulardesign. These security principles promote self-sufficient components that would be more easily replaced in the event of damage or security vulnerabilities. 
            \begin{tcolorbox} [width=\linewidth, sharp corners=all, colback=white!90!green]
            \paragraph{Key Alignment:} Compartmentalized designs are more secure \textit{and} are easier to reuse and recycle.
            \end{tcolorbox}
                
            Generally, the idea of automatically removing unnecessary data (described in the security principle \remnantremoval) facilitates a sustainable EoL for components across all three sustainability themes by eliminating security risks arising from sensitive data remaining on systems that are reused or recycled. Moreover, the automated removal of data simplifies a system's EoL, as owners might not need to wipe the system manually.
            \begin{tcolorbox} [width=\linewidth, sharp corners=all, colback=white!90!green]
            \paragraph{Key Alignment:} Automated data clean-up reduces security risks \textit{and} encourages sustainable disposal.
            \end{tcolorbox}

        \subsubsection{Tension in the Disposal and End of Life Phase}
            Our analysis highlighted that disposal, of both the physical system and the data in it, can put sustainability and security in tension. In a physical sense, improperly disposing of systems (i.e., not recycled) generates e-waste. However, disposing of systems without removing sensitive information could allow attackers to retrieve the data via ``dumpster diving'' attacks~\cite{garfinkel2003remembrance,sutherland2010zombie,camp_how_2023}. It is also important to consider the digital disposal of information on systems. Accumulation of data on a system can reduce the performance of certain operations and might present a security risk if attackers can leverage accumulated data to execute more effective attacks. Moreover, users may not remember all the information that has digitally accumulated on their systems, which risks their information leaking when disposing of the system. These risks might encourage someone to destroy systems before disposing of them to avoid breach of their privacy. From a sustainability perspective, this is undesirable as the system might be feasibly reused by other owners, or could have valuable materials recycled. Security or sustainability consequences are more likely when only one is considered in isolation, but the risks can be minimized by carefully considering both the sustainability and security impacts of decisions.
            
            \begin{tcolorbox} [width=\linewidth, sharp corners=all, colback=white!90!red]
            \paragraph{Key Tension:} Sustainable disposal strategies can introduce security risks if naively implemented. 
            \end{tcolorbox}

    \subsection{Overarching Relationships}
        Despite a trend towards an alignment of themes and principles in Tables~\ref{tab:designdev}, \ref{tab:usage}, and \ref{tab:disposal}, it would be an oversimplification to claim that security and sustainability are exclusively in alignment or tension. 
        There are cases where security or sustainability decisions can benefit the other, and this is most likely when the decision is made with consideration of potential consequences on both security and sustainability.
        Designers and developers can mitigate possible negative impacts on sustainability when making decisions that prioritize security through thoughtful implementation and compromise. Below, we expand on our identified points of alignment, points of tension, and where there is no relationship between security principles and sustainability themes.

        \paragraph{Separation of Security and Sustainability:}
            Three of the security principles, \datatypevalidation, \trustanchorjustification, and \requestresponseintegrity, seem disconnected from sustainability. These principles deal with more granular security concerns than the other security principles. As a mirror to this, granular sustainability principles may not intersect with security (e.g., sustainability guidelines about leveraging renewable energy sources for manufacturing; the source of the energy does not have a significant impact on cybersecurity). The claims from previous work that security and sustainability must be addressed separately~\cite {kocksch2023investigating, bradleyEscaping2023, keleherBalancing2025} are not fundamentally incorrect, but, as described below, do not capture the nuanced relationship of security and sustainability.

        \paragraph{Overarching Alignment:}
            The majority of non-blank cells in Tables~\ref{tab:designdev}, \ref{tab:usage}, and \ref{tab:disposal} indicate positive relationships between the sustainability themes and security principles. We believe that it is possible to apply sustainability or security guidelines in ways that lead to both better sustainability \textit{and} security. 

            In particular, the security principles \simplicityandnecessity, \isolatedcompartments, and \modulardesign advocate for design practices that would achieve the outcomes described in the sustainability themes \simplifyandreduce, \adaptandaugment, and \repairandmaintain. In security, it is easier to address vulnerabilities in simple, compartmentalized systems. It is also beneficial to minimize potential points of attack for bad actors on such systems, as sensitive information can be shared only with components that need it. For sustainability, simple systems are less wasteful and easier to maintain for long lifespans. Moreover, it is easier to replace components from compartmentalized systems, and such systems are easier to augment with new modules or components. System maintainability is a goal of both security and sustainability; therefore, making a system more maintainable is mutually beneficial.

            We saw significant overlap in terms of prioritizing users. In both security and sustainability, unnecessary friction can push users towards behaviors with negative impacts. For example, confusing security alerts might cause users to disregard legitimate risks or to dispose of a system that merely requires a software update. Moreover, key aspects of extending system lifespans (e.g., maintenance, repair, and reuse) require user buy-in, so poor usability can be an obstacle to system longevity. Usable security strategies reduce friction caused by security mechanisms, which would, in turn, eliminate a potential aspect of the system discouraging long-term use. To encourage sustainable behaviors, it is important to align with users' mental models, just as usable security seeks to design systems where the secure path is the easiest one. Overall, these alignments indicate not only the positive influence of one domain on the other but also suggest that security and sustainability can be mutually reinforcing when practitioners carefully consider both dimensions during system design.

        \paragraph{Overarching Tension:}
            We noticed that three security principles \completemediation, \sufficientworkfactor, and \reluctantallocation were more likely than others to have tension with sustainability themes. These principles reflect a more closed-off approach to security. While they describe practices that legitimately improve security, they should be used judiciously as they may necessitate undesirable trade-offs. 
            In fact, these principles even seem to have tension with other security principles. For example, closed systems that require the user to grant permission every time they interact with external agents are likely to cause frustration, thereby violating the \userbuyin security principle. Rather than indicating a fundamental contradiction between pursuing sustainability and security, these points of tension reflect the complex nature of security in real-world use cases and suggest that careful designs can mitigate the consequences on security or sustainability. 

            \evidenceproduction was another security principle that tended to have tension with the sustainability themes. In this case, the tension stems from creating too much or too detailed log data that might reveal sensitive info if left on the system during disposal. A naive outcome may encourage the destruction of the system rather than recycling or reuse, but incorporating proper EoL security features would more effectively mitigate these risks and facilitate sustainable recycling or reuse practices. 

    \subsection{Limitations}
        Our analysis is influenced by our selected set of security principles; while we believe these to be a representative set, other relationships might appear if a similar study were conducted with other security principles. Similarly, our qualitative analysis relies on subjective interpretations by our research team. 
        Through iterative processes and discussion among team members, we believe that our interpretations are reasonable. 
        We acknowledge that others may identify different individual relationships, but believe that our overarching argument holds: security and sustainability have many points of alignment. Lastly, we have focused our arguments on the sustainability and security relationships, but other external pressures, such as economic incentives, may also practically drive decisions that put the two domains in opposition. 

    \section{Key Takeaways} \label{dis:users}
        A significant contributor to the difficulty of security is that there is no end state or goal where security has been achieved. While the security principles offer valuable guidance for improving the security of a system, the application of security principles does not guarantee that the product or system is perfectly secure. Attaining sustainability faces similar challenges. It is impossible to account for all issues systems might encounter in real-world use that might limit their lifespan. Moreover, even applying individual sustainability practices may involve trade-offs. For example, systems or products might be overbuilt for typical conditions; while this reduces the risk of failure and possibly increases longevity, it also indicates a waste of resources in the design and development. A similar phenomenon exists in security; increasing password complexity to avoid brute-force attacks increases the likelihood that users will find workarounds to remember the passwords, and this poses new risks (e.g., writing down the password near the computer, or reusing passwords across multiple accounts).
        For both security and sustainability, experience or contextual knowledge would be necessary to pre-filter guidelines to use only the most relevant for a given use case. Designers and developers should keep all the guidelines in mind during the design and development, and revisit them throughout the product life cycle. 

        \paragraph{Takeaway 1: The User is Not the Weakest Link in Security \textit{or} Sustainability.}
            Usable security researchers have worked to combat the myth that users are the weakest link in terms of security (e.g.,~\cite{hielscherEmployeesWhoDont}). 
            The myth relies on the notion that a user's inability to follow security best practices is a personal failure rather than a systemic issue with the guidance or design of the system. 
            \citet{Maniates2002} flags that a similar false blame regarding sustainability has been foisted onto end users and claims that regulators and manufacturers are abdicating their own responsibility. 
            When disposing of a system, the user is responsible for sending it to a landfill or making a more sustainable choice (e.g., recycling the system or finding a way for it to be reused). However, placing this blame solely on a user ignores the reality of the constraints they must act within. Was a system designed to last? A user can only be expected to use a system so long as it remains functional. If a system arbitrarily loses functionality, and there has been no misuse of the system, then the manufacturer or designer is to blame for the short system lifespan. A system might not be designed such that it is recyclable, or a user might not live in a place that facilitates the recycling of e-waste. Moreover, poor data security design may impede users from deleting, or confirming the deletion of, personal data from a system before they transfer ownership or dispose of it. In fact, a lack of tools for sanitizing devices might limit users from desired behaviors such as reuse. Fortunately, the guidelines we found in our review tended not to place blame on users and recognized that design influences the user's behavior. 
        
        \paragraph{Takeaway 2: Following computer security design principles is more likely to benefit than harm sustainability goals.}
            As indicated in Tables~\ref{tab:designdev}, \ref{tab:usage}, and \ref{tab:disposal}, the adherence to security principles will likely result in products or systems that also meet several of the sustainability themes (or at least remain neutral with respect to sustainability, as noted by the abundance of empty cells). Whereas past work has explicitly or implicitly presented sustainability and security as two concerns that must be individually addressed and balanced~\cite{bradleyEscaping2023,kocksch2023investigating,keleherBalancing2025}, we find that careful application of either security or sustainability guidelines simultaneously benefits both.
            Broadly speaking, security principles that are more process-focused (e.g., \opendesign or \modulardesign) tend to be more in alignment with sustainability themes than those that are more granular or specific (e.g., \evidenceproduction or \requestresponseintegrity). Our analysis found that only four of the 20 surveyed security principles are in (limited) tension with sustainability goals.

        \paragraph{Takeaway 3: Dogmatic application of guidelines drives tension between security and sustainability.}
            We observe points of tension when either security or sustainability is hyper-prioritized. This tension is magnified in security principles that could be interpreted as restrictive or centralizing. For example, the \completemediation principle calls for verifications of access rights before a resource is accessed, and this is typically enforced through some centralized authority. While central authorization allows for consistent security policy management and enforcement, it introduces a single point of failure that may render the system unusable and lead to e-waste if no recovery mechanism is considered. Thoughtful design can account for these failure modes, ensuring the inclusion of fallback paths that preserve device functionality even when individual components are unavailable.  

        \paragraph{Takeaway 4: Unlike sustainability, security ignores product end of life and beyond.}
            There is a history of advocacy for security to be incorporated into the digital design and development process from the beginning~\cite{microsoft2026security, cisa2023secure}, rather than ``bolted on'' to production systems. However, the emphasis on ``secure by design'' throughout design and development all but disappears at the end of a product's life cycle. The most secure means of deletion is through physical destruction of storage media, but this is an unsustainable practice if the storage media has a remaining usable lifespan. Beyond basic guidance on data privacy and deletion, and factory reset options, security and sustainability remain unreconciled at EoL. 
            On one hand, security decisions might prevent sustainable practices relating to EoL. For example, a system might be designed to prevent ownership transfer to avoid exposing the original owner's data.
            On the other hand, decisions relating to the EoL of a system may present security risks. Without a usable and effective means of deleting personal data from a system, a user puts their personal information at risk when transferring ownership of the system, with obvious negative security consequences. Alternatively, a user may choose not to recycle or reuse their system if they cannot reliably remove their data, leading to negative sustainability consequences. 

    \section{Real-world Examples of Security Affecting Sustainability} \label{sec.examples}
     Using our sustainability themes and security principles, we analyze three real-world examples and illustrate the complex interactions between security and sustainability in practice.

        \paragraph{Google Chromecast Certificate Expiration:} 
            In 2024, Google ended production of the 10-year-old second-generation Chromecast~\cite{ChromecastUsersReport2025}. While Google was supporting the product, the company followed appropriate security principles, preventing (to the best our knowledge) wide-scale security issues with Chromecast systems. However, Google seemingly did not implement plans to renew the 10-year root certificate store when they decided to deprecate the product, causing user systems to malfunction after failing certificate validation checks when contacting Google servers (another application of \safedefaults), effectively turning every affected Chromecast into e-waste. After public pressure, Google issued a firmware update that migrated affected systems to a new certificate authority with an expiration date set for 2045~\cite{thomson2025chromecast}, but this fix merely delays the same problem by 20 years rather than addressing the underlying design failure: Google did not plan for the system's cryptographic stack to outlive the product's support period. 

        \paragraph{Microsoft Windows 11:} 
            Microsoft broke backwards compatibility with older hardware by requiring a Trusted Platform Module (TPM) 2.0 for Windows 11~\cite{brankWhatTPMWhy2025}. TPM 2.0 implements \completemediation by verifying the integrity of each stage of the boot process, ensuring that only trusted firmware and operating system code is executed. However, by pairing this requirement to Windows 11 and simultaneously ending support for Windows 10, Microsoft is artificially ending the useful life of millions of older PCs that lack this hardware. This is a failure to consider \degrade: rather than offering a degraded but functional (i.e., without TPM) security model on older hardware that still provides meaningful security protections, Microsoft offers users an all-or-nothing choice.
            We note that Microsoft has the technical capacity to allow running Windows 11 without a TPM. Registry keys such as \texttt{BypassTPMCheck} and \texttt{BypassSecureBootCheck} allow installation on unsupported hardware, signalling that this is an artificial constraint rather than a technical impossibility. A more sustainable approach would be for Microsoft to support a reduced-security configuration of Windows 11 for older hardware, clearly communicating the security trade-offs to users rather than forcing them to choose between an unsupported operating system and discarding functional hardware.

        \paragraph{Android Bootloader Security:}
            Android's locked bootloader serves a legitimate security purpose for the primary device owner. It prevents unauthorized firmware from being loaded, ensures verified boot integrity, and protects against data extraction from lost or stolen devices~\cite{redini2017bootloader}. However, once a manufacturer ends official software support, a locked bootloader prevents the device owner from installing community-maintained firmware that could extend the device's useful life (e.g., through LineageOS~\cite{lineageos}). While vendors such as Samsung have removed the bootloader unlocking option from some of its devices~\cite{salvogiangri2025bootloader}, others, such as Fairphone, explicitly treat bootloader unlocking as a core feature and actively support the custom Android OS community long after official support ends. This difference illustrates that bootloader security can be used as a tool for forcing hardware obsolescence rather than for its intended use: protecting the system while the user wants that protection.

\section{Research Agenda} \label{sec:agenda}
    The relationship between security and sustainability is an emerging research area with many opportunities for further investigation. Our analysis suggests that the assumed incompatibility between these domains is largely a myth, but dismantling that myth requires empirical studies, new tools, and more actionable guidance. Given our identified parallels with usable security, particularly around usability and the individualization of responsibility, we believe the usable security community is well-positioned to lead much of this work. 

    \paragraph{1. Empirical Studies of Security-driven E-Waste.}
    We have presented several high-profile cases where security decisions contributed to premature device disposal, but there is little empirical data on the prevalence of this pattern. Part of the challenge is that the boundary between security updates and feature updates is often unclear. When a user discards a device that no longer receives software updates, it is difficult to determine whether the primary motivating factor was loss of security support, loss of new features, degraded performance, or some combination. Understanding the role that security plays in disposal decisions requires the same types of empirical methods the usable security community already employs to study user behavior; longitudinal studies, surveys, and interviews that examine how users reason about security and obsolescence in their everyday lives.

    \paragraph{2. Measuring Security and Sustainability.} Measurements and metrics are a common struggle for both sustainability and security. Carbon dioxide emissions are a popular metric for assessing environmental impact, but this metric fails to account for other aspects of sustainability, such as system repairability or recyclability. New metrics are needed to capture the relationship between security decisions and device longevity. For example, mean time to disposal and mean number of unique owners might offer insights into how effectively a device or system was designed for long-term ownership or reuse. Dedicated research is needed to develop and validate these metrics across diverse device categories and usage contexts.
    \paragraph{3. Security at End of Life.} Our analysis of security principles and real-world use cases identified a lack of security guidance when a product's support period ends (Section~\ref{dis:users}, Takeaway 4). While the privacy community has engaged with secure data deletion~\cite{reardon2013sok,ceci2021concerned}, there is little guidance for the broader set of EoL challenges, such as secure ownership transfer, migration to community-maintained support, or responsibly winding down security support. These challenges can introduce their own security concerns. Publicly releasing firmware source code makes it easier for community maintainers to extend a device's life, but it makes it easier for attackers to discover vulnerabilities that will never receive an official fix. Community-maintained software can extend device lifespans, but users must trust the maintainers. The usable security community has addressed similar human factors problems for years, and we hope this expertise can be drawn into sustainable security as well. 

    \paragraph{4. Sustainable Security Analysis Tools.} In the field of security, threat modeling tools (e.g., attack trees and STRIDE~\cite{shostack2014threat}) help designers and developers systematically identify possible attack vectors to design against. We found no equivalent tools for reasoning about threats to sustainability, especially when considering security-induced sustainability failures. Developing such tools, potentially by adapting existing security analysis methods, could help designers anticipate and mitigate sustainability risks early in the design process.

    \paragraph{5. Developing Actionable Guidance.} Design guidelines do not always give developers specific, actionable guidance. Our analysis suggests this impacts both sustainability and security: high-level principles are broadly applicable but require significant effort to operationalize, which may lead to them being overlooked. Sustainability is a complex, context-dependent issue, so it may be more productive to develop processes for understanding contexts (e.g., probing questions, analysis tools) rather than providing blanket guidelines. Given the shared challenges related to end users and developers, it is particularly important to involve the usable security community in this effort. 
    While there are efforts to develop regulatory solutions~\cite{higginbothamLongevity2025}, they do not provide clear instructions for designers or developers. One promising direction is the development of sustainable security design patterns: validated solutions to common problems that bridge both fields~\cite{buschmann1996pattern}. Our real-world examples in Section~\ref{sec.examples} suggest that corresponding anti-patterns could also be defined to flag common pitfalls. However, developing such guidance will require collaboration between security and sustainability experts, and this collaboration would benefit from accessible shared language for topics ranging from cryptography to materials recycling.

\section{Conclusion}
    We conducted a systematization of knowledge to determine the relationship between cybersecurity and environmental sustainability, synthesizing 12 sustainability themes from \numpapers papers in the sustainable design literature. We compared these themes to a broad set of security design principles and found that following security design principles is more likely to benefit than harm sustainability goals. The few points of tension stem from a narrow application of individual principles rather than fundamental incompatibility between domains. Similar to security, sustainability suffers from the myth that users are the weakest link.
    Overall, the security and sustainability alignments indicate not only the positive influence of one domain on the other but also suggest that security and sustainability can be mutually reinforcing when practitioners carefully consider both dimensions during system design. We believe that the usable security community is uniquely positioned to advance security without sacrificing sustainability.

\balance
\bibliographystyle{plainnat}
\bibliography{2026_sustainsec_SoK_bib}

\clearpage
\appendix
\section{Computer Security Design Principles} \label{app:secprinciples}
\small
The following principles originally appeared in Chapter 1 of \textit{Computer Security and the Internet: Tools and Jewels from Malware to Bitcoin} \cite{van2021computer}. The principles (excluding chapter cross-references and author notes) are reproduced below with permission of the author and copyright holder for convenience. 

\subsection*{Two High-Level Principles:}

\paragraph{\securitybydesign.} Build security in, starting at the initial design stage of a development cycle, since
secure design often requires core architectural support 
absent if security is a late-stage add-on.
Explicitly state the \textit{design goals}
of security mechanisms and what they are \textit{not} 
designed to do, since it is impossible to evaluate effectiveness without knowing goals.
In design and analysis documents, explicitly state all security-related \textit{assumptions}, 
especially related to trust and trusted parties (supporting P17);
note that a security policy itself might not specify assumptions. 

\paragraph{\designforevolution.} Design base architectures, mechanisms, and protocols to support evolution,
including \techemph{algorithm agility} for graceful upgrades of
crypto algorithms (e.g., encryption, hashing) with minimal 
impact on related components. 
Support automated \techemph{secure software update} where possible.
Regularly re-evaluate the effectiveness of security mechanisms,
in light of evolving threats, technology, and architectures; be ready to
update designs and products as needed. 
\vspace{10pt}

\subsection*{Twenty Principles:}

\noindent\textbf{\simplicityandnecessity.} Keep designs  
as simple and small as possible. 
Reduce the number of components used, retaining only those that are essential; minimize functionality,
favor minimal installs,   
and disable unused functionality. 
Economy and frugality in design---especially for core security 
mechanisms---simplifies analysis and reduces errors and oversights.  
Configure initial deployments to have non-essential  
services and applications disabled by default (related to P2). 

\noindent\textbf{\safedefaults.} Use safe default settings (beware, defaults often go unchanged).
For access control, deny-by-default. 
Design services to be \techemph{fail-safe}, here meaning that when they fail,
they fail ``closed" (e.g., denying access).
As a method, use explicit inclusion via
\techemph{allowlists} (\textit{goodlists}) of authorized entities with all others denied, 
rather than exclusion via \techemph{denylists} (\textit{badlists}) that would allow all those unlisted.

\noindent\textbf{\opendesign.} Do not rely on secret designs, attacker ignorance, or \techemph{security by obscurity}.
Invite and encourage open review and analysis. 
For example, the \textit{Advanced Encryption Standard} was selected from a set of public candidates by open review;
undisclosed cryptographic algorithms are now widely discouraged.
Without contradicting this, 
leverage unpredictability where advantageous, as
arbitrarily  publicizing tactical defense details   
is rarely beneficial
(there is no gain in advertising to thieves that you are on vacation, or posting
house blueprints); and
beware exposing error messages or timing data that vary based on secret values, lest they leak the secrets.
 
\noindent\textbf{\completemediation.} For each access to every object, and ideally immediately before the access is to be granted,
verify proper authority. Verifying authorization requires authentication (corroboration of an identity),
checking that the associated principal is authorized, and checking that the request
has integrity (it must not be modified after being issued by the legitimate party---cf. P19). 

\noindent\textbf{\isolatedcompartments.} Compartmentalize system components using strong isolation structures (containers) that manage or prevent 
cross-component communication, information leakage, and control.
This limits damage when failures occur, and protects against
\techemph{escalation of privileges}; P6 and P7
have similar motivations.
Restrict authorized cross-component communication to observable
paths with defined interfaces to aid mediation, screening, and use of \techemph{chokepoints}.
Examples of containment means include:
process and memory isolation, disk partitions,   
virtualization, software guards, zones, gateways and firewalls. 

\noindent\textbf{\leastprivilege.} Allocate  
the fewest privileges needed for a task, 
and for the shortest duration necessary.
For example, retain superuser privileges
only for actions requiring them; drop and reacquire privileges if needed later. Do not use
a Unix \textit{root} account for tasks where regular user privileges suffice. 
This reduces exposure, and limits damage from the unexpected.
P6 complements P5 and P7.

\noindent\textbf{\modulardesign.} Avoid monolithic   
designs that embed full privileges into large single components;    
favor object-oriented and finer-grained designs 
that segregate privileges (including address spaces) across smaller units or 
processes.
P6 guides more on the \textit{use} of  
privilege frameworks, P7 more on designing base architectures. 

\noindent\textbf{\smalltrustedbases.} Strive for small code size in components that must be  
trusted, i.e., components on which a larger system strongly depends for security.
For example, high-assurance systems centralize critical security services 
in a minimal core operating system \techemph{microkernel}, 
whose smaller size allows efficient concentration of security analysis. 

\noindent\textbf{\timetestedtools.} Rely wherever possible on time-tested, expert-built security tools including
protocols, cryptographic primitives and toolkits,   
rather than designing and implementing your own. 
History shows that security design and implementation is difficult to get right even for experts; 
thus amateurs are heavily
discouraged  
(\textit{don't reinvent a weaker wheel}).
Confidence increases with the length of time mechanisms and tools
have survived under load (sometimes called \textit{soak testing}).

\noindent\textbf{\leastsurprise.} Design mechanisms, and their user interfaces, to behave as users expect. 
Align designs with users' mental models of their protection goals, to reduce user mistakes that compromise security.
Especially where errors are irreversible
(e.g., sending confidential data or secrets to outside parties), tailor to the experience of target users;
beware designs suited to experts but triggering mistakes by ordinary users. 
Simpler, easier-to-use (i.e., \techemph{usable}) mechanisms yield fewer surprises.

\noindent\textbf{\userbuyin.} Design security mechanisms that users are motivated to use rather than bypass,
and so that users' path of least resistance is a safe path. 
Seek design choices that illuminate benefits, improve user experience, and minimize inconvenience.
Mechanisms viewed as time-consuming, inconvenient  or 
without perceived benefit risk non-compliance. 
Example: a subset of Google gmail users  voluntarily use  a two-step authentication scheme, which augments basic passwords   
by one-time passcodes sent to the user's phone.

\noindent\textbf{\sufficientworkfactor.} For configurable security mechanisms where the probability of attack success
increases predictably with effort, tune the mechanism so that
the cost to defeat it (\techemph{work factor}) clearly exceeds the resources of anticipated classes of adversaries. Thus
use suitably strong defenses.
Example 1: random cryptographic keys should be sufficiently long to preclude being found by brute-force search
(i.e., enumerative guessing attacks).
Example 2: disallow user-chosen passwords that are so weak that a feasible number of 
guesses yields a non-negligible chance of success.

\noindent\textbf{\defenseindepth.} Build defenses in multiple layers backing each other up,    
forcing attackers to defeat independent layers, thereby avoiding \techemph{single points of failure}. 
If any layer relies on alternative defense segments, design each
to be comparably strong (``equal-height fences" for \textit{defence-in-breadth})  
and strengthen the weakest segment first
(smart attackers jump the lowest bar or break the \techemph{weakest link}). 
As a design assumption, expect  
some defenses to fail on their own due to errors, and that attackers will defeat 
others more easily than anticipated, or entirely bypass them. 

\noindent\textbf{\evidenceproduction.} Record system activities   
through event logs, monitoring tools, and other means to promote accountability,
help understand and recover from system failures,
and support intrusion detection tools. 
Example:
robust \textit{audit trails} support \techemph{forensic analysis} tools, to   
help reconstruct events 
related to intrusions and criminal activities. 
In many cases, mechanisms that facilitate detection and evidence production
may be more cost-effective than outright attack prevention. 

\noindent\textbf{\datatypevalidation.} Verify that all received data meets expected (assumed) properties or data types. 
If data input is expected, ensure that it cannot be processed as code by subsequent components.
This may involve \techemph{canonicalization} of data
(e.g., of fragmented packets or encoded URL characters)
and broader \techemph{input sanitization} 
(e.g., to address \techemph{code injection} or \techemph{command injection} attacks.

\noindent\textbf{\remnantremoval.} On termination of a session or program, remove all traces of sensitive data 
associated with a task,   
including secret keys and any
remnants recoverable from secondary storage, RAM and cache memory.
Note that common file deletion removes directory entries, whereas
\techemph{secure deletion} aims to make file content unrecoverable even by forensic tools.
Related to remnant removal, beware that  
while a process is active, information may leak elsewhere by \techemph{side channels}.

\noindent\textbf{\trustanchorjustification.} Ensure or justify confidence   
placed in any base point of assumed trust,   
especially when mechanisms 
iteratively or transitively extend trust from a base point
(such as a \techemph{trust anchor}   
in a browser \techemph{certificate chain}). 
More generally, verify trust assumptions 
where possible,
with extra diligence at registration, initialization, software installation, and starting points in the lifecycle
of a software application, security key or credential. 

\noindent\textbf{\idependentconfirmation.} Use simple, independent (e.g., local device) cross-checks to increase confidence   
in code or data, especially if it may arrive from outside domains or over untrusted channels.
Example: integrity of downloaded software applications or public keys 
can be confirmed by comparing a locally computed \techemph{cryptographic hash} of the item to a ``known-good" hash obtained over  
an independent channel (voice call, text message, widely trusted site). 

\noindent\textbf{\requestresponseintegrity.} Verify that responses match  
requests in \techemph{name resolution} and other
distributed protocols. Their design should   
include cryptographic integrity checks that 
bind steps to each other within a given transaction or protocol run
to detect unrelated or substituted responses;  
beware protocols lacking authentication. 
Example: a \techemph{certificate request} specifying a unique 
subject name or domain expects in response a certificate for that subject; this field in the
response certificate should be cross-checked against the request.

\noindent\textbf{\reluctantallocation.} Be reluctant to allocate resources or expend effort
in interactions with unauthenticated, external agents. 
For services open to all parties,  
design to mitigate intentional resource consumption. 
Place a higher burden of  
effort on agents that initiate an interaction.
(A party initiating a phone call should not be the one to demand:  \textit{Who are you?} When possible, authenticate.
Related: processes should beware abuse as a conduit extending their privileges to unverified agents.)

\section{Papers Used in the Analysis}

\begingroup
\renewcommand*{\arraystretch}{1.2}
    \begin{table*}[tb]
        \centering
          \rowcolors{1}{white}{gray!20}
        \scriptsize
        \begin{tabular}{lp{2.2in}p{3.5in}c}
        \toprule
            \textbf{Year} & \textbf{Authors} & \textbf{Title }& \textbf{Citation }\\ \hline
            2018 & Mourão, Brunna C.; Karita, Leila; Do Carmo Machado, Ivan & Green and Sustainable Software Engineering - a Systematic Mapping Study & \cite{mouraoGreenSustainableSoftware2018a} \\ 
            
            2021 & Hansson, Lon Åke Erni Johannes; Cerratto Pargman, Teresa; Pargman, Daniel Sapiens & A Decade of Sustainable HCI: Connecting SHCI to the Sustainable Development Goals & \cite{hanssonDecadeSustainableHCI2021a} \\ 
            
            2022 & Bremer, Christina; Knowles, Bran; Friday, Adrian & Have We Taken On Too Much?: A Critical Review of the Sustainable HCI Landscape & \cite{bremerHaveWeTaken2022a} \\ 
            
            2007 & Mankoff, Jennifer C.; Blevis, Eli; Borning, Alan; Friedman, Batya; Fussell, Susan R.; Hasbrouck, Jay; Woodruff, Allison; Sengers, Phoebe & Environmental sustainability and interaction & \cite{mankoffEnvironmental2007} \\ 
            
            2007 & Blevis, Eli & Sustainable interaction design: invention \& disposal, renewal \& reuse & \cite{blevisSustainableInteractionDesign2007a} \\ 
            
            2016 & Cohn, Marisa Leavitt & Convivial Decay: Entangled Lifetimes in a Geriatric Infrastructure & \cite{cohnConvivial2016} \\ 
            
            2014 & Penzenstadler, Birgit; Raturi, Ankita; Richardson, Debra; Tomlinson, Bill & Safety, Security, Now Sustainability: The Nonfunctional Requirement for the 21st Century & \cite{penzenstadlerSafety2014} \\ 
            \bottomrule
        \end{tabular}
        \caption{Round 1 Seed papers used for forwards-backwards review}
        \label{tab:seedpapers}
    \end{table*}
\endgroup

\begingroup
\renewcommand*{\arraystretch}{1.2}
    \begin{table*}[tb]
              \rowcolors{1}{white}{gray!20}
        \centering
        \scriptsize
        \begin{tabular}{lp{1.85in}p{3.5in}cc}
            \toprule
            \textbf{Year} & \textbf{Authors }& \textbf{Title} & \textbf{Citation} & \textbf{\# Guidelines}  \\ \hline
            
            2016 & Chen, Jay & A strategy for limits-aware computing & \cite{chenStrategyLimitsawareComputing2016} & 4  \\ 
            
            2015 & Penzenstadler, Birgit & A Toolkit for SE for Sustainability - A Design Fiction & \cite{penzenstadlerToolkitSESustainability2015} & 4  \\ 
            
            2006 & Blevis, Eli & Advancing Sustainable Interaction Design: Two Perspectives on Material Effects & \cite{blevisAdvancingSustainableInteraction2006} & 10  \\ 
            
            2013 & Håkansson, Maria; Sengers, Phoebe & Beyond being green: simple living families and ICT & \cite{hakanssonBeingGreenSimple2013} & 4  \\ 
            
            2015 & Remy, Christian; Gegenbauer, Silke; Huang, Elaine M. & Bridging the Theory-Practice Gap: Lessons and Challenges of Applying the Attachment Framework for Sustainable HCI Design & \cite{remyBridgingTheoryPracticeGap2015} & 7  \\ 
            
            2002 & Jain, Ravi; Wullert, John & Challenges: environmental design for pervasive computing systems & \cite{jainChallengesEnvironmentalDesign2002} & 6  \\ 
            
            2012 & Tomlinson, Bill; Silberman, M. Six; Patterson, Donald; Pan, Yue; Blevis, Eli & Collapse informatics: augmenting the sustainability & \cite{tomlinsonCollapseInformaticsAugmenting2012} & 6  \\ 
            
            2003 & Abraham, Martin A.; Nguyen, Nhan & Green Engineering: Defining the Principles - Results from the Sandestin Conference. & \cite{andersonCommunicatingSustainability2008} & 9  \\ 
            
            2010 & Jung, Heekyoung; Blevis, Eli; Stolterman, Erik & Conceptualizations of the Materiality of Digital Artifacts and their Implications for Sustainable Interaction Design & \cite{jung2010conceptualizations} & 5  \\ 
            
            2018 & Nyström, Tobias; Mustaquim, Moyen & Exploring Sustainable HCI Research Dimensions Through the Inclusive Innovation Framework & \cite{nystromExploringSustainableHCI2018} & 4  \\ 
            
            2023 & Khalifeh, Amin; Al-Adwan, Ahmad Samed; Alrousan, Mohammed Kasem; Yaseen, Husam; Mathani, Belal; Wahsheh, Firas Rashed & Exploring the Nexus of Sustainability and Project Success: A Proposed Framework for the Software Sector & \cite{khalifehExploringNexusSustainability2023} & 8  \\ 
            
            2010 & Huh, Jina; Nam, Kevin; Sharma, Nikhil & Finding the lost treasure: understanding reuse of used computing devices & \cite{huhFindingLostTreasure2010} & 2  \\ 
            
            2017 & Blevis, Eli; Preist, Chris; Schien, Daniel; Ho, Priscilla & Further Connecting Sustainable Interaction Design with Sustainable Digital Infrastructure Design & \cite{blevisFurtherConnectingSustainable2017} & 4  \\ 
            
            2014 & Niemelä, Marketta; Ikonen, Veikko; Leikas, Jaana; Kantola, Kristiina; Kulju, Minna; Tammela, Antti; Ylikauppila, Mari & Human-Driven Design: A Human-Driven Approach to the Design of Technology & \cite{niemelaHDD20214} & 3  \\ 
            
            2025 & Englhardt, Zachary; Hähnlein, Felix; Mei, Yuxuan; Lin, Tong; Sun, Connor Masahiro; Zhang, Zhihan; Patel, Shwetak; Schulz, Adriana; Iyer, Vikram & Incorporating Sustainability in Electronics Design: Obstacles and Opportunities & \cite{englhardtIncorporatingSustainabilityElectronics2025} & 6  \\ 
            
            2020 & Khalifeh, Amin; Farrell, Peter; Alrousan, Mohammad; Alwardat, Shaima; Faisal, Masar & Incorporating sustainability into software projects: a conceptual framework & \cite{khalifehIncorporatingSustainabilitySoftware2020} & 8  \\ 
            
            2012 & Gegenbauer, Silke; Huang, Elaine M. & Inspiring the design of longer-lived electronics through an understanding of personal attachment & \cite{gegenbauerInspiringDesignLongerlived2012} & 3  \\ 
            
            2008 & Odom, William & Personal inventories: toward durable human-product relationships & \cite{odomPersonalInventoriesDurable2008} & 3  \\ 
            
            2011 & Pierce, James; Paulos, Eric & Second-hand interactions: investigating reacquisition and dispossession practices around domestic objects & \cite{pierceSecondhandInteractionsInvestigating2011} & 4  \\ 
            
            2020 & Pham, Yen Dieu; Bouraffa, Abir; Maalej, Walid & ShapeRE: Towards a Multi-Dimensional Representation for Requirements of Sustainable Software & \cite{phamShapeREMultiDimensionalRepresentation2020} & 9  \\ 
            
            2012 & Hazas, Mike; Brush, A. J. Bernheim; Scott, James & Sustainability \textit{does not} begin with the individual & \cite{hazasSustainabilityDoesNot2012} & 3  \\ 
            
            2015 & Becker, Christoph; Chitchyan, Ruzanna; Duboc, Leticia; Easterbrook, Steve; Penzenstadler, Birgit; Seyff, Norbert; Venters, Colin C. & Sustainability Design and Software: The Karlskrona Manifesto & \cite{beckerSustainabilityDesignSoftware2015} & 9  \\ 
            
            2007 & Blevis, Eli & Sustainable interaction design: invention \& disposal, renewal \& reuse & \cite{blevisSustainableInteractionDesign2007a} & 5  \\ 
            
            2017 & Remy, Christian; Bates, Oliver; Thomas, Vanessa; Huang, Elaine M. & The Limits of Evaluating Sustainability & \cite{remyLimitsEvaluatingSustainability2017} & 4  \\ 
            
            2011 & Burger, Paul; Christen, Marius & Towards a capability approach of sustainability & \cite{burgerCapabilityApproachSustainability2011} & 4  \\ 
            
            2023 & Cheng, Tingyu; Abowd, Gregory D.; Oh, Hyunjoo; Hester, Josiah & Transient Internet of Things: Redesigning the Lifetime of Electronics for a More Sustainable Networked Environment & \cite{chengTransientInternetThings2023} & 4  \\ 
            
            2016 & Preist, Chris; Schien, Daniel; Blevis, Eli & Understanding and Mitigating the Effects of Device and Cloud Service Design Decisions on the Environmental Footprint of Digital Infrastructure & \cite{preistUnderstandingMitigatingEffects2016} & 5  \\ 
            
            2024 & Lu, Jasmine; Lopes, Pedro & Unmaking Electronic Waste & \cite{luUnmakingElectronicWaste2024} & 3  \\ 
            
            2024 & Kilic, Damla; Sailaja, Neelima & User-Centred Repair: From Current Practices to Future Design & \cite{kilicUserCentredRepairCurrent2024} & 5 \\ 
        \bottomrule
        \end{tabular}
        \caption{The \numpapers papers included for analysis. The \textit{\# Guidelines} column indicates the number of guidelines extracted per paper.}
        \label{tab:corpus}
    \end{table*}
\endgroup

\end{document}